\documentclass[prb,showpacs,twocolumn]{revtex4-1}
\usepackage{amsmath,amssymb}
\usepackage[final]{graphicx}

\newcommand{\+}{\circlearrowright}
\renewcommand{\-}{\circlearrowleft}

\begin{document}

\title{Interplay between uniaxial strain and magnetophonon resonance in graphene}

\author{Oleksiy Kashuba}
\affiliation{Institut f\"ur Theorie der Statistischen Physik, RWTH Aachen, 52056 Aachen, Germany}
\email{kashuba@physik.rwth-aachen.de}
\author{Vladimir I. Fal'ko}
\affiliation{Department of Physics, Lancaster University, Lancaster, LA1~4YB, UK}

\begin{abstract}
We study the fine structure of the phonon $G$ peak in the inelastic light scattering spectra of strained graphene in the presence of a quantizing magnetic field.
We show that under the conditions of the magnetophonon resonance (MPR), the spectral line shape in Raman spectra of the $\Gamma$-point longitudinal optic--transverse optical phonon undergoes multiple splitting and acquires a strong dependence on the polarization of the incoming and detected light.
Interference of strain and MPR leads to the modification of the anticrossing branches, introducing the variation of the light polarization preferences depending on the proximity to the resonance, and their better resolution between neighboring anticrossings.
\end{abstract}

\pacs{78.67.Wj, 63.22.Rc, 78.20.Ls, 71.70.Di}
% 78.67.Wj Optical properties of graphene
% 63.22.Rc Phonons in graphene
% 78.20.Ls Magneto-optical effects
% 71.70.Di Level splitting and interactions: Landau levels

\maketitle

%\section{Introduction}

One of the most pronounced features in Raman spectra of graphene is a peak corresponding to the excitation of $\Gamma$-point longitudinal optic--transverse optical (LO-TO) phonons, known as a ``$G$ peak''.\cite{Ferrari2006}
Electron-phonon interaction affects the phonon dispersion and decay\cite{GeimNovoselov2007,Pinczuk2007,Ando2006,CastroNeto2007}, as well as couples these phonon modes to the light\cite{Basko2008-9}.
The renormalization of the LO-TO phonon by coupling with electrons is most pronounced in the presence of a quantizing magnetic field which leads to the discrete spectrum of Landau levels (LLs) in graphene\cite{Ando2007}.
When energies of some inter-Landau-level transitions resonate with the optical phonon energy, which is a condition known as magnetophonon resonance, the electron-phonon ($e$-ph) interaction leads to the anticrossing line shape of the phonon mode in the Raman spectrum\cite{Ando2007,Kechedzhi2007,Kashuba2012,Potemski2011,Smirnov2013}.
Besides the use of Raman spectroscopy for the analysis of the $e$-ph coupling in graphene, Raman measurements of phonon excitation can be used to probe strain in graphene\cite{RMPgraphene,Guinea2009g,Sasaki2010}.
Splitting of the $G$ peak by strain has already been observed experimentally\cite{Ni2008,Huang2009,Mohiuddin2009,Galiotis2010,Cheong2012}, in agreement with density functional theory (DFT) modeling of the Gr\"uneisen parameter\cite{Thomsen2009,Cheng2011}.

The renormalization of the optical phonon spectrum by the $e$-ph interaction is caused by a mutual transformation of the phonon into a virtual electron-hole pair and back.
In a zero magnetic field this leads to the corrections to the dispersion relation of optical phonons and widening of the $G$ peak\cite{Ando2006}.
In the presence of a strong magnetic field the spectrum of electrons is discrete, with Landau levels $n^{\pm}$ at $\epsilon_{n^{\pm}}=\pm\frac{\hbar v}{\lambda_{B}}\sqrt{2n}$ ($v$ is a Dirac velocity in graphene, and $\lambda_B=\sqrt{\hbar c/e B}$ is a magnetic length) in the conduction ($+$) and valence ($-$) band, and a peculiar zero-energy LL ($n=0$) at the band edge.
The discrete electron spectrum makes coupling between the lattice mode and electron-hole excitations resonant when the optical phonon energy $\omega_{G}=1586\,{\rm cm}^{-1}=197\,{\rm meV}$ matches the energy $\Omega_{n}=\sqrt{2}\frac{\hbar v}{\lambda_{B}}(\sqrt{n+1}+\sqrt{n})$ of an $e$-$h$ excitation with an orbital momentum $\Delta n=\pm1$ and belonging to the representation $E_{2}$ of the $C_{6v}$ symmetry group of the graphene lattice\cite{Kashuba2012,Kechedzhi2007,Ando2007}.
As a result, magnetophonon resonance (MPR) involves a pair of interband transitions $n^{-}\to(n+1)^{+}$ (with $\Delta n=+1$) and $(n+1)^{-}\to n^{+}$ (with $\Delta n=-1$) with energies $\Omega_{n}$, and each of these two electronic modes is coupled to the phonons with opposite circular polarizations.\cite{Note1}
Here we stress that the MPR in graphene has a built-in natural selectivity of coupling with a circularly polarized (angular momentum $\Delta n=\pm1$) optical lattice vibration, which is enhanced when one of the two resonant electronic modes is switched of by Pauli blocking in heavily doped graphene.
In contrast to MPR, strain transforms the degenerate $\Gamma$-point LO-TO optical phonon modes into a pair of phonons with energies $\omega_{G^{+}}$ and $\omega_{G^{-}}$, linearly polarized along the two perpendicular axes of the strain tensor and split proportionally to the amount of strain, $\varepsilon$.

In this Rapid Communication we study the interplay between the fine structures in the optical phonon spectrum induced by strain and the MPR, and predict its manifestation in magneto-Raman spectra of graphene.
Since both of these effects are sensitive to the polarization of the lattice mode probed by Raman spectroscopy, the resulting MPR in strained graphene acquires a peculiar polarization and carrier density dependent form, which is analyzed in details for two major MPR conditions: $\Gamma$-point optical phonon resonance with $0\to1^{+}/1^{-}\to0$ (at $B\sim25\,{\rm T}$) and $1^{-}\to2^{+}/2^{-}\to1^{+}$ (at $B\sim4.5\,{\rm T}$) inter-Landau-level transitions.

%\section{Raman spectral density}

The $\Gamma$-point optical phonon in graphene is characterized by the relative displacement $\mathbf{u}=\frac{1}{\sqrt{2}}(\mathbf{u}_{A}-\mathbf{u}_{B})$ of atoms in sublattices $A$ and $B$.
Together, the LO-TO lattice modes, the electronic excitations and their coupling are described using the Hamiltonian
\begin{align}
\nonumber
&H=H_\text{ph}+H_\text{e}+H_\text{$e$-ph},\quad H_\text{e}=\!\!\!\sum_{n^{\alpha}=0,1^{\pm},\ldots}\!\!\!\epsilon_{n^{\alpha}} c_{n^{\alpha}}^{+} c_{n^{\alpha}},
\\
&H_\text{ph}=\frac{M}{2}(\dot{\mathbf{u}}^{2}+\omega_{0}^{2}\mathbf{u}^{2}) + 2M\omega_{0}w\sum_{\upsilon,\nu=x,y} u_{\upsilon}\varepsilon_{\upsilon\nu}u_{\nu},
\\\nonumber
&H_\text{$e$-ph}=-g\sqrt{2M\omega_{0}}
\!\!\!\sum_{\genfrac{}{}{0pt}{}{n^{\alpha},m^{\beta}=0,1^{\pm},\ldots}{n-m=\pm1}}\!\!\!
(\mathbf{s}_{n^{\alpha}m^{\beta}}\cdot\mathbf{u}) c_{n^{\alpha}}^{+} c_{m^{\beta}}.
\end{align}
Here, $\omega_{0}$ is a ``bare'' phonon frequency, formally defined before the $e$-ph interaction is taken into account, and $M$ is a reduced mass of carbon atoms.
The phonon mode splitting by strain is determined by a strain tensor,
\begin{gather}
\label{eq:strain}
\hat{\varepsilon}=
\begin{pmatrix}
\varepsilon_{xx} &  \varepsilon_{xy} \\
\varepsilon_{xy} & -\varepsilon_{xx}
\end{pmatrix}
=
R_{\theta}^{-1}
\begin{pmatrix}
\varepsilon &  0 \\
0 & -\varepsilon
\end{pmatrix}
R_{\theta},
\quad
\mathrm{Tr}\,\hat\varepsilon=0,
\\\nonumber
\varepsilon_{xx}=\varepsilon\cos2\theta,
\quad
\varepsilon_{xy}=\varepsilon\sin2\theta,
\end{gather}
and angle $\theta$ determines the rotation $R_{\theta}$ of the principal axes of the strain tensor with respect to the crystalline axes (zigzag direction) of the graphene lattice.\cite{Note2}
The size of this splitting is characterized by the parameter $w\equiv\frac{1}{2}\partial(\omega_{G^{-}}-\omega_{G^{+}})/\partial\varepsilon\approx 13\,{\rm cm}^{-1}\!/\%$ as estimated in Ref.~\onlinecite{Galiotis2010}.
The electron-phonon interaction is parametrized by the coupling constant $g\approx0.2\,{\rm eV}$~\cite{Ferrari2006}.
The selection rules for electron-phonon interaction in $H_\text{$e$-ph}$ are incorporated into matrix elements $\mathbf{s}_{(n+1)^{\alpha}n^{\beta}}= (\mathbf{s}_{n^{\beta}(n+1)^{\alpha}})^{*}$, such that $\mathbf{s}_{(n+1)^{\alpha}n^{\beta}} = \frac{1}{\sqrt{2}}\alpha\mathbf{e}_{\+}$ for $n\geq1$ and $\mathbf{s}_{1^{\alpha}0} = \alpha\mathbf{e}_{\+}$, where $\mathbf{e}_{\+/\-}=\frac{1}{\sqrt{2}}(\mathbf{e}_{x} \pm i \mathbf{e}_{y})$ are  unit vectors which determine circularly polarized modes\cite{Ferrari2006,Kashuba2012}.

To describe the renormalization of the phonon spectrum and its manifestation in Raman inelastic scattering, we estimate the full retarded Green's function of a phonon,
\begin{equation}
D_{\upsilon\nu}(\omega)=2m\omega_{0}\int_{0}^{\infty}\langle[u_{\upsilon}(t),u_{\nu}(0)]\rangle e^{i(\omega+i0)t}dt,
\end{equation}
where square and angle brackets denote the commutation with subsequent averaging over the ground state.
Using a random phase approximation (RPA)\cite{Kashuba2012} the latter can be described as
\begin{equation}
\begin{cases}
D^{-1}_{\upsilon\nu}&=
\left(D^{(0)}_{\upsilon\nu}\right)^{-1}-\frac{2}{\sqrt{3}}\frac{g^{2}}{\gamma_{0}^{2}}\Pi_{\upsilon\nu}-w\varepsilon_{\upsilon\nu},
\\
D^{(0)}_{\upsilon\nu}&=
\delta_{\upsilon\nu}\Bigr/(\omega-\omega_{0}+i\gamma_{G}).
\end{cases}
\end{equation}
Here, $\gamma_{G}$ stands for phonon relaxation rate, $\gamma_{0}\approx2.7\,{\rm eV}$ is the $A$--$B$ hopping parameter for graphene\cite{RMPgraphene}, related within a tight-binding approximation to the electron velocity as $\gamma_{0}=\frac{2}{3}\hbar v/a_\text{C--C}$ ($a_\text{C--C}$ is the distance between two neighbor carbon atoms in a graphene lattice), and
\begin{multline}
\Pi_{\upsilon\nu}(\omega) =(v\hbar)^{2} \rho_{B} \sum_{n^\alpha m^\beta}\int \frac{d\epsilon}{2\pi} 
s^{\upsilon}_{m^\beta n^\alpha} s^{\nu}_{n^\alpha m^\beta} 
\\
\frac{-i}{2}\Bigl\{ G^{A}_{n^\alpha}(\epsilon) G^{K}_{m^\beta}(\epsilon+\omega) + G^{K}_{n^\alpha}(\epsilon) G^{R}_{m^\beta}(\epsilon+\omega) \Bigr\}
\label{eq:POgeneral}
\end{multline}
is a retarded polarization operator taken for given electron density $n_e$ and magnetic field $B$ which determines $\rho_{B}=2/\pi\lambda_{B}^{2}$ density of electron states in a single LL (spin and valley degeneracies are taken into account).
Electronic Green's functions are
\begin{align*}
&G^{R/A}_{n^\alpha}(\epsilon)=[\epsilon-\epsilon_{n^\alpha}\pm i0]^{-1},
\\
&G^{K}_{n^\alpha}(\epsilon)=[1-2f(\epsilon)][G^{R}_{n^\alpha}(\epsilon)-G^{A}_{n^\alpha}(\epsilon)],
\end{align*}
where $f(\epsilon)$ is a Fermi function.
In principle, the sum over the electronic states in Eq.~\eqref{eq:POgeneral} is accumulated over the entire Brillouin zone.
For the sake of the following analysis, this contribution can be separated into 
\begin{equation*}
\Pi_{\upsilon\nu}(\omega)=\left.\Pi_{\upsilon\nu}(0)\right|_{B=0,n_e=0}+\Delta\Pi_{\upsilon\nu}(\omega),
\end{equation*}
where the value $\left.\Pi_{\upsilon\nu}(0)\right|_{B=0,n_e=0}$ of the polarization operator for undoped graphene in zero magnetic field at zero frequency can be used to determine the renormalized phonon frequency\cite{Ando2006,CastroNeto2007},
\begin{equation*}
\omega_{G}=\omega_{0}-\tfrac{2}{\sqrt{3}}\tfrac{g^{2}}{\gamma_{0}^{2}}\Pi_{\nu\nu}(0)\bigr|_{B=0,n_e=0}.
\end{equation*}
The latter can be calculated\cite{Kechedzhi2007,Kashuba2012} using the tight-binding model, while all the features of low-energy electron states near the Dirac point in a magnetic field, doping, and the frequency dependence are incorporated in
\begin{subequations}
\begin{align}
\Delta&\Pi_{\upsilon\nu} =
(\mathbf{e}_{\+})_{\upsilon}(\mathbf{e}_{\+})_{\nu}^{*}\Delta\Pi_{\+}
+
(\mathbf{e}_{\-})_{\upsilon}(\mathbf{e}_{\-})_{\nu}^{*}\Delta\Pi_{\-},
\\
\Delta&\Pi_{\+}(\omega)=\bigl[\Delta\Pi_{\-}(-\omega)\bigr]^{*}=
\nonumber\\&=
\frac{\Omega_{0}^{2}}{2\pi}
\sum_{n=0}^{\infty} \Biggl[
\frac{2}{\Omega_{n}} +
\sum_{\pm}\frac{f(\epsilon_{(n+1)^{\pm}})-f(\epsilon_{n^{\mp}})}{\omega\pm\Omega_{n}}+
\nonumber\\&\hspace{2.3cm}+
\sum_{\pm}\frac{f(\epsilon_{(n+1)^{\pm}})-f(\epsilon_{n^{\pm}})}{\omega\pm\Omega_{n}'}
\Biggr].
\end{align}
\label{eq:POexplicit}
\end{subequations}
where $\Omega_{n}=\sqrt{2}\frac{\hbar v}{\lambda_{B}}(\sqrt{n+1}+\sqrt{n})$ accounts for interband transitions, $\Omega_{n}'=\sqrt{2}\frac{\hbar v}{\lambda_{B}}(\sqrt{n+1}-\sqrt{n})$ accounts for intraband transitions, and spin and valley degeneracy is taken into account.
The term corresponding to the first transition $1^{\pm}\leftrightarrow0$ comes with an effective factor 2, which is due to the difference of LL wave functions for $n=0$ and $n\geq1$.

Using group symmetry analysis, one can show that the optical $\Gamma$ phonon, which belongs to the $E_{2}$ representation of the $C_{6v}$ symmetry group, couples to a pair of photons through the vectors $\mathbf{d}$, determined in terms of polarization vectors $\mathbf{l}$ ($\tilde{\mathbf{l}}$) of an absorbed (emitted) photon and transforming according to the same irreducible representation $E_{2}$~\cite{Basko2008-9,Kashuba2012},
\begin{equation}
I=-\frac{1}{2\pi}\mathrm{Im}\sum_{\upsilon,\nu=x,y} d^{*}_{\upsilon}D_{\upsilon\nu}d_{\nu},
\quad
\mathbf{d}=
\begin{pmatrix}
l_{x}\tilde{l}^{*}_{y} + l_{y}\tilde{l}^{*}_{x} \\
l_{x}\tilde{l}^{*}_{x} - l_{y}\tilde{l}^{*}_{y}
\end{pmatrix}.
\end{equation}
Then, the spectral density of Raman scattering (normalized by the total Raman efficiency of the $G$ peak) is determined as
\begin{align}
\label{eq:IGgeneral}
I=&\frac{1}{2\pi}
\mathrm{Im}\left[\frac{1}{\det A} \sum_{\xi,\eta=\+,\-} (\mathbf{d}\mathbf{e}_{\xi})(\mathbf{d}\mathbf{e}_{\eta})^{*}A_{\xi\eta}\right];
\\\nonumber
&\det A = A_{\+\+}A_{\-\-}-w^{2}\varepsilon^{2},
\\\nonumber
&A_{\+\+(\-\-)} = \omega_{G} + \lambda\Delta\Pi_{\-(\+)} - \omega - i\gamma_{G},
\\\nonumber
&A_{\-\+} = A_{\+\-}^{*} = w(\varepsilon_{xx}+i\varepsilon_{xy}).
\end{align}
Using $(\mathbf{d}\mathbf{e}_{\+/\-}) = \pm i\sqrt{2}(\mathbf{l}\mathbf{e}_{\-/\+})(\tilde{\mathbf{l}}\mathbf{e}_{\+/\-})^{*}$, we find that for circularly polarized right ($\+$) and left ($\-$) incoming and outgoing photons
\begin{equation}
I_{\+\rightleftarrows\-}=\frac{1}{\pi}\mathrm{Im}\frac{A_{\-\-/\+\+}}{\det A},
\quad
I_{\+\to\+(\-\to\-)}=0.
\end{equation}
Using definitions in Eq.~\eqref{eq:strain}, we find that for linearly polarized photons,
\begin{equation}
I_{\varphi\to\tilde{\varphi}}\!=\!\frac{1}{\pi}\mathrm{Im}\frac{\tfrac{1}{2}(A_{\+\+}\!+\!A_{\-\-}) \!+\! w \epsilon \cos(2\varphi\!+\!2\tilde{\varphi}\!+\!2\theta)}{\det A},
\end{equation}
where angles $\varphi$ and $\tilde\varphi$ determine the direction of the polarization vector of incoming ($\mathbf{l}$) and outgoing ($\tilde{\mathbf{l}}$) photons with respect to the crystalline axis corresponding to the zigzag direction in graphene.
Finally, for non-polarized incoming light the total spectral density for the polarization-integrated Raman spectrum is
\begin{equation*}
\langle I\rangle = \frac{1}{2\pi}\mathrm{Im}\frac{A_{\+\+}+A_{\-\-}}{\det A}.
\end{equation*}

%\section{Results}

\begin{figure}
\centering
\includegraphics[height=10.5cm]{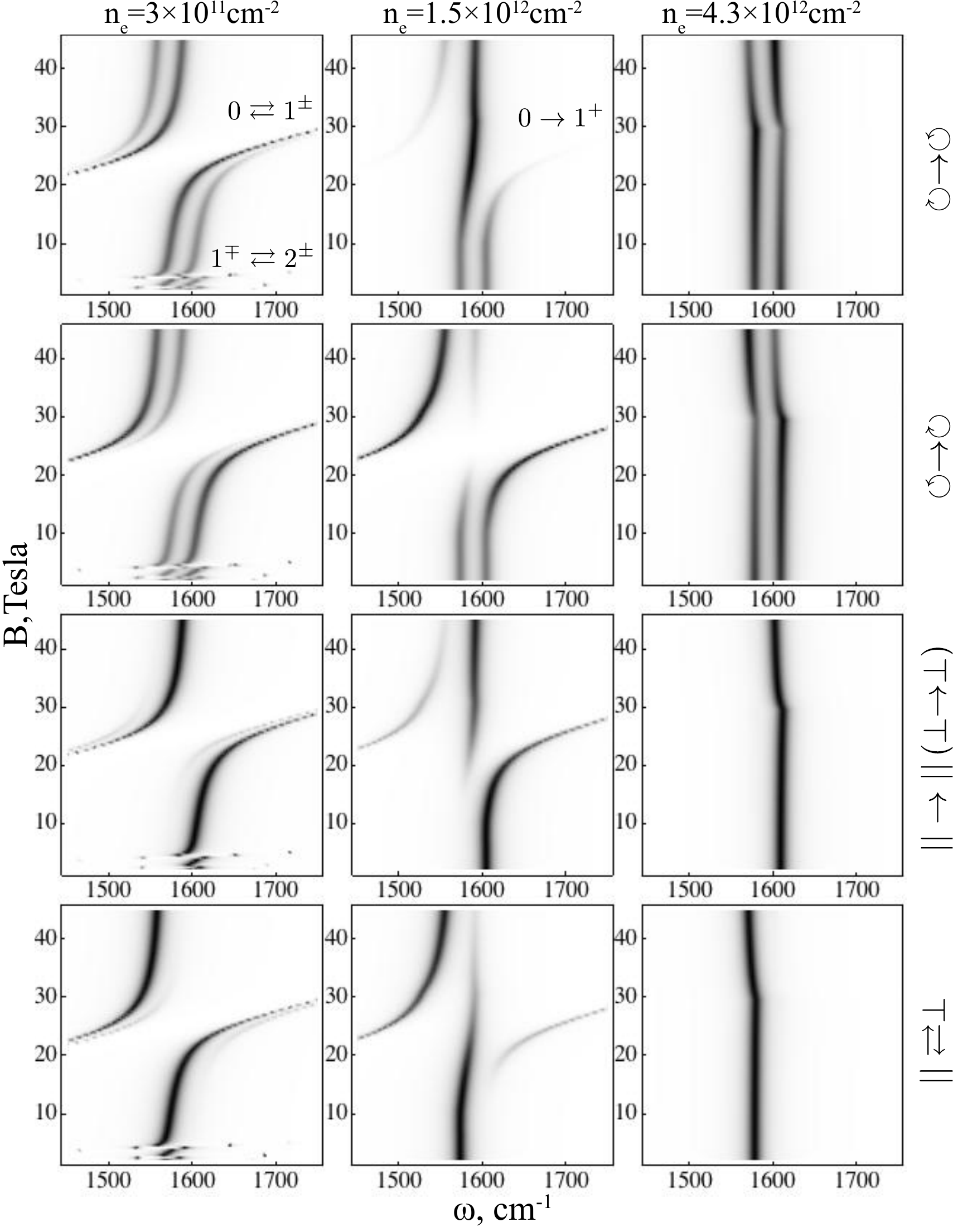}
\caption{The intensity of the Raman spectrum in graphene strained along the zigzag axis as a function of magnetic field $B$.
These spectra are plotted for a phonon relaxation rate $\gamma_{G}=6\,{\rm cm}^{-1}$ and strain $\varepsilon=1.2\%$.
Three columns correspond to different electron densities, and rows to different polarizations.
Symbols $\circlearrowright(\circlearrowleft)$ stand for right- (left-) handed circular polarization of the incident and scattered light; $||$ and $\perp$ denote linear polarization of light parallel and perpendicular to the principal elongation axis of the strain tensor.
}
\label{fig:plots1}
\end{figure}

\begin{figure}
\centering
\includegraphics[height=10.5cm]{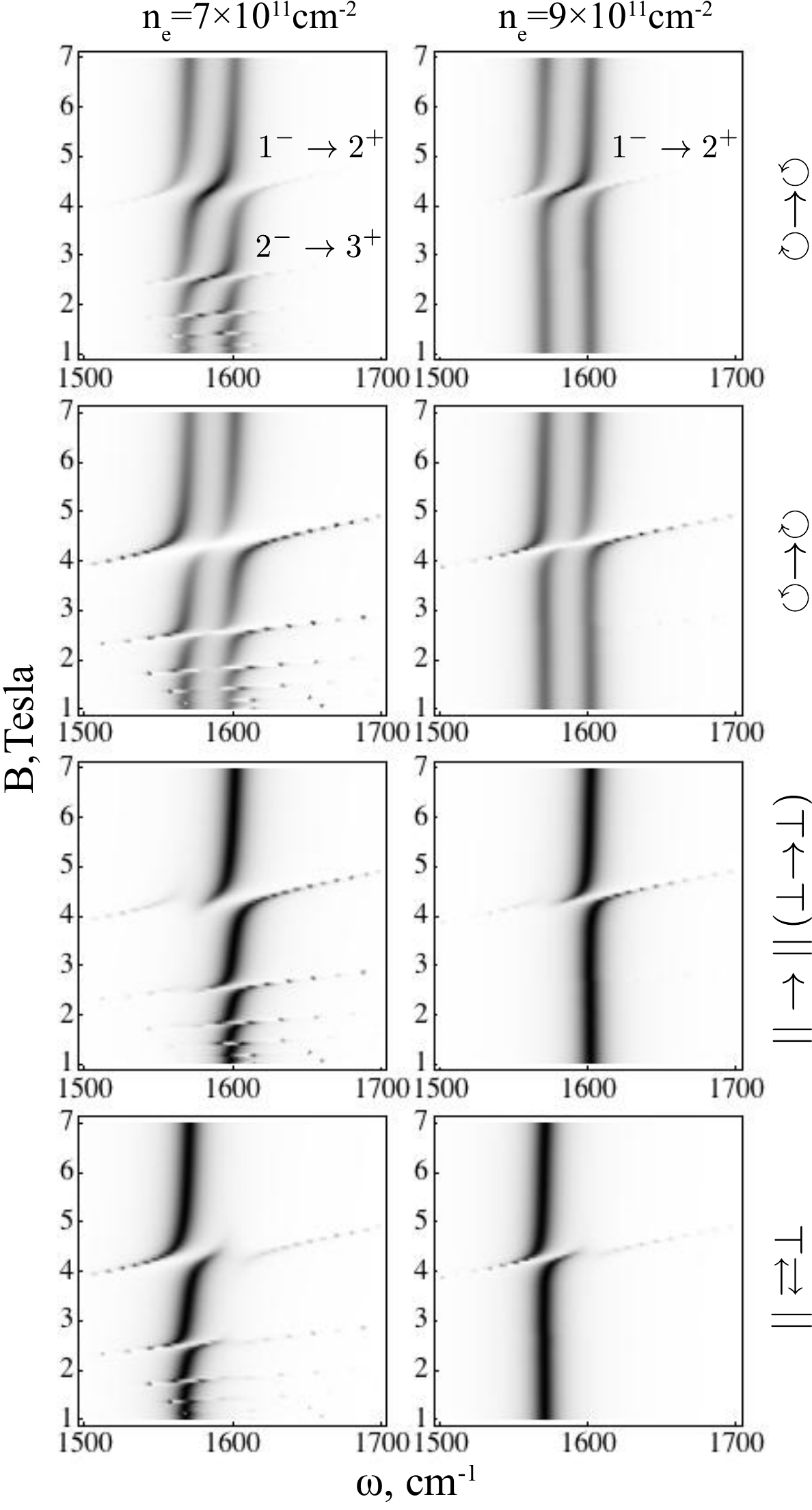}
\caption{Low magnetic field spectrum for strain $\varepsilon=1.2\%$ along the zigzag axis, phonon relaxation rate $\gamma_{G}=6\,{\rm cm}^{-1}$.
Here, anticrossing $2^{-}\to 1^{+}$ is blocked, and $1^{-}\to 2^{+}$ is active in Raman spectra.
}
\label{fig:plots2}
\end{figure}

The resulting LO-TO phonon spectral line shape formed by the interplay between the MPR (which promotes anticrossing of circularly polarized lattice modes with the inter-LL excitations) and strain (which promotes splitting of two linearly polarized phonons into independent Raman signals) is shown in Figs.~\ref{fig:plots1} and~\ref{fig:plots2}.
The value of strain $\varepsilon=1.2\%$ used in these plots is such that the resulting phonon splitting $w|\varepsilon|$ and the MPR anticrossing energy $g$ have comparable sizes.
To illustrate the MPR-strain interplay, we plot the calculated grayscale maps of the phonon line intensity in polarization-resolved Raman spectra for in and out photons with circular and linear polarizations, which interrogate differently polarized lattice modes.
The interplay between strain and MPR is most pronounced when the inter-LL transition $(n+1)^{-}\to n^{+}$ is forbidden by Pauli blocking of the Landau level $n^{+}$, while the transition to the Landau level $(n+1)^{+}$ is possible, which takes place in medium-doped graphene (middle column in Fig.~\ref{fig:plots1}).
For doping $n_{e}=1.5\times 10^{12}\,{\rm cm}^{-2}$ and magnetic field $B\approx27\,{\rm T}$ which corresponds to the main $\omega_{G}\approx\sqrt{2}\hbar v/\lambda_{B}$ MPR condition, the zero-energy LL is full, and the Landau level $1^{+}$ is only partially filled.
Thus, in the vicinity of the anticrossing, only one of two circularly polarized phonon modes undergoes resonant mixing with the LL transition $0\to1^{+}$, whereas the other remains unperturbed (middle column in the top two rows in Fig.~\ref{fig:plots1}).
Away from the resonance, the phonon line is split into two linearly polarized branches, which is reflected by the linear-polarization-selective line shape in the bottom rows of Fig.~\ref{fig:plots1}.
Note that for the density $n_{e}=1.5\times10^{12}\,{\rm cm}^{-2}$, all anticrossings at lower magnetic fields are blocked by electrons filling completely the relevant LLs, whereas for the higher density (right-hand column in Fig.~\ref{fig:plots1}) even the MPR with the $0\to1^{+}$ transition is blocked.
One may also notice a kink in the position of the phonon line in the latter case, resulting from a partial depletion of the level $1^{+}$, which activates the coupling of the $0\to1^{+}$ mode with the phonon, downshifting the energy of the latter.
Finally, the MPR anticrossing in low-density or undoped graphene, shown in the left-hand column, displays an independent anticrossing behavior of two linearly polarized phonons separately.
This happens because couplings of phonons with $0\to1^{+}$ and $1^{-}\to0$ transitions are almost equal, and this degeneracy allows two linearly polarized phonons to mix with equal strength with two linearly polarized interband transitions.

Figure~\ref{fig:plots2} demonstrates the higher ($n^{-}\to n^{+}+1$ and $n^{-}+1\to n^{+}$, $n\geqslant1$) LL anticrossings at smaller magnetic fields.
The abundance of those can only be seen in the lower-density structures, where the Fermi level for electrons is low enough, $E_{F}<\omega_{G}/2$.
The calculated spectra for this case are shown in the left-hand column of Fig.~\ref{fig:plots2}, where the MPR displays the independent anticrossing behavior of two split linearly polarized phonons, which is characteristic for the low-density structures.
The size of strain is also important in determining the MPR line shape.
For weak strain, $w|\varepsilon|\ll g$, the lineshape and density dependence of MPR is essentially the same as in non-strained structures analyzed in Ref.~\onlinecite{Kechedzhi2007}.
For the dominating strain, $w|\varepsilon|\gg g$, each of the two linearly polarized phonons undergoes anticrossing with a linear combination of $n^{-}\to (n+1)^{+}$ and $(n+1)^{-}\to n^{+}$ transitions, resulting in the MPR fine structure displayed in Fig.~\ref{fig:plots2}.

Besides the direct influence of strain on MPR via changing the properties of LO-TO phonons, one may consider the effect of static lattice deformations on MPR, via the modification of electronic properties of graphene in the nonlinear response to strain\cite{Vozmediano2010,Vozmediano2012} and generation of a valley-asymmetric pseudo-magnetic field by inhomogeneous lattice deformations\cite{RMPgraphene}.
Out of those two, the influence of strain on Dirac velocity, making it slightly anisotropic, in linear approximation can be merely reduced to the weak renormalization of the magnetic field.\cite{Note3}
The role of inhomogeneous strain on the inter-LL transitions may be much more drastic, leading to inhomogeneous broadening and an additional structure in the MPR line shape, as discussed in detail in Refs.~\onlinecite{Kashuba2012,Smirnov2013}.

%\begin{acknowledgments}
The authors thank A.~Ferrari and M.~Potemski for useful discussions.
This work was supported by ERC Advanced Grant ``Graphene and Beyond'', EPSRC Science and Innovation Award, and the Royal Society Wolfson Research Merit Award.
%\end{acknowledgments}

%\bibliographystyle{revtexsimple}
%\bibliography{mprs}

\end{document}